\documentstyle[aps,twocolumn,floats,epsfig]{revtex}

\begin{document}


\title{Quasiclassical Theory of Spontaneous Currents at Surfaces and
Interfaces of $d$-Wave Superconductors}
\author{M.H.S. Amin$^1$, A.N. Omelyanchouk$^2$,  S.N. Rashkeev$^3$,
M. Coury$^1$, and A.M. Zagoskin$^{1,4}$}
\address{$^1$D-Wave Systems Inc., 320-1985 West Broadway, Vancouver,
B.C., V6J 4Y3, Canada; $^2$B.I. Verkin Institute for Low
Temperature Physics and Engineering, Ukrainian National Academy
of Sciences, Lenin Ave. 47, Kharkov 310164, Ukraine;  $^3$Dept.
of Physics and Astronomy, Vanderbilt University, Box 1807 Station
B, Nashville, TN 37235, USA; $^4$Physics and Astronomy Dept., The
University of British Columbia, 6224 Agricultural Rd., Vancouver,
B.C., V6T 1Z1, Canada}

\address{~
\parbox{14cm}{\rm
\medskip
We investigate the properties of spontaneous currents generated at
surfaces and interfaces of $d$-wave superconductors using the
self-consistent quasiclassical Eilenberger equations. The
influence of the roughness and reflectivity of the boundaries on
the spontaneous current are studied. We show that these have very
different effects at the surfaces compared to the interfaces,
which reflects the different nature of the time reversal symmetry
breaking states in these two systems. We find a signature of the
``anomalous proximity effect'' at rough $d$-wave interfaces. We
also show that the existence of a subdominant order parameter,
which is necessary for time reversal symmetry breaking at the
surface, suppresses the spontaneous current generation due to
proximity effect at the interface between two superconductors. We
associate orbital moments to the spontaneous currents to explain
the ``superscreening'' effect, which seems to be present at all
ideal $d$-wave surfaces and interfaces, where $d_{xy}$ is the
favorite subdominant symmetry. }} \maketitle

%

\section{Introduction}

The qualitative difference between $d$-wave and conventional
($s$-wave) superconductors is the intrinsic $\pi$-shift of the
order parameter phase between the crystallographic $a$ and $b$
directions \cite{SigristRice92}. The most remarkable effects due
to it are realized near surfaces and interfaces of such
superconductors. In particular, depending on the direction of the
crystalline axes on both sides of a Josephson junction, the
equilibrium can be achieved not only at phase difference $\phi
=0$, like in conventional Josephson junctions, but at $\phi =\pi
$ as well (``$\pi$-junctions'')
\cite{SigristRice92,Geshkenbein87}. Therefore, a frustrated ring
(a ring with an odd number of $\pi $-junctions) will have a time
reversal symmetry breaking ground state, which supports a
spontaneous magnetic flux $\pm \Phi _{0}/2$, where $\Phi_0=\pi/e$
is the flux quantum (throughout this article we take
$\hbar=k_B=1$). Experiments \cite{Wollman93,VH95,Tsuei96}
confirmed this prediction, thus establishing $d_{x^2-y^2}$
pairing symmetry of high-$T_{c}$ cuprates.

The $\pi $-junctions are not specific to unconventional orbital
symmetry (being first predicted\cite{Geshkenbein87} and recently
realized \cite{Ryazanov00} in SFS structures). They do not either
provide the only way that $\mathcal{T}$-breaking states can
appear in $d$-wave superconductors. We can distinguish between two
situations: a doubly connected geometry, like a ring with $\pi
$-junctions, and a simply connected one, like a single boundary
between a $d$-wave superconductor and another differently
oriented $d$-wave superconductor, $s$-wave superconductor, or
vacuum. In the former case, the spontaneous flux is quantized (in
units of $\Phi _{0}/2$) \cite {SigristRice92}. Therefore, the
spontaneous currents are always present and flow through the
whole ring (within the screening length from the surface which is,
in high-$T_{c}$ cuprates, of order 1500 \AA ). In the latter case,
on the other hand, there is no quantization condition for the
spontaneous flux and it can take arbitrary values \cite{Huck,ZO}.
The spontaneous currents can be absent or confined to a much
narrower area near the surface/interface itself, as we will see
later.

A natural description of $\mathcal{T}$-breaking states near the
surfaces and interfaces of $d$-wave superconductors can be given
in the language of Andreev bound states \cite{Andreev}. They are
formed by off-diagonal scattering of quasiparticles by a spatially
inhomogeneous pairing potential, $\Delta({\bf r})$. Off-diagonal
means that the reflected quasiparticle changes the branch of the
dispersion law (particle to hole and vice versa), so that
electric charge $2e$ is being transferred to or from the
condensate. Remarkably, Andreev reflection conserves spin
(exactly) and momentum (with accuracy $E/E_{F}$, $E$ being the
quasiparticle energy measured from the Fermi level, $E_F$). As a
result, the reflected hole is sent back with almost the same
group velocity as the incident electron, and will therefore
retrace its path (up to distance $\sim \xi _{E}= v_{F}/E$) in the
clean limit (large elastic scattering length).

Mathematically, the Schr\"{o}dinger equation for the (quasi-)
electron wave function is now replaced by a matrix
Bogoliubov--de~Gennes equation \cite{Zagoskin} for two-component
quasiparticle (bogolon) wave function $\Psi({\bf r}) = \left(
_{v({\bf r})}^{u({\bf r})}\right) $. The order parameter $\Delta
({\bf r})$ and its complex conjugate play the role of the
off-diagonal components of the matrix scattering potential. If we
neglect the self-consistency condition, which expresses $\Delta
({\bf r})$ through $u({\bf r}),v({\bf r})$, the equations are
easily solved, and the positions of Andreev levels, $E$, are
found from the Bohr-Sommerfeld quantization condition for a
quasiclassical trajectory bounded by Andreev reflections at
points $L,R$ \cite{Zagoskin}:
\begin{equation}
\int_{L}^{R}p_{_{\rm elec}}(E)dl-\int_{R}^{L}p_{_{\rm
hole}}(E)dl+\phi -\beta (E)=2\pi n.  \label{BSquant}
\end{equation}
Here, the first and second terms represent the phase gain along
the trajectory, $\phi $ is the phase difference between the order
parameters in the points $L,R$ (including the intrinsic phase
difference), and the additional phase shift $\beta (E)=\pi \cdot
O(1)$ depends on the shape of the scattering potential $\Delta
({\bf r})$ near the reflection points ($\beta (0)=\pi $). The
minus sign before the second term reflects the quasiparticle
branch change. Evidently, the right-moving electron and
left-moving hole both carry electric current in the same
direction. Therefore, unlike conventional standing wave solutions
to the Schr\"{o}dinger equation, Andreev bound states (Andreev
levels)(Eq.\ (\ref{BSquant})) carry electric current. This
provides a mechanism for the Josephson effect in such structures
as microbridges or in SNS contacts \cite{Kulik69,Ishii,BJ72},
which can be extended to the case of a general Josephson junction
(see \cite{FurusakiReview}). The current through a junction of
unit width is expressed as \cite{RiedelBagwell}

\begin{figure}[t]
\epsfysize 5cm \epsfbox[-50 220 300 650]{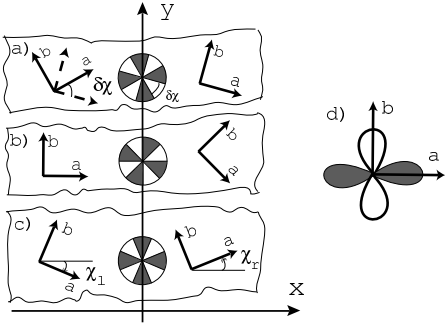} \caption{Grain
boundary junction between two $d$-wave superconductors. Here $a$
and $b$ denote the crystalline axes, and $\protect\delta \chi $ is
the mismatch angle. The $y$-axis is chosen along the grain
boundary. The diagrams in the middle indicate the directions
connecting the lobes of the order parameter with the same sign
(shaded) and opposite sign, on both sides of the grain boundary.
Supercurrent in the shaded directions is carried by regular
Andreev levels, otherwise by $\pi$-levels (see text). For given
$\delta\chi$, the properties of the junction depend on the grain
boundary orientation vs. axes $a,b$. The special cases are
asymmetric $(\chi_l=0)$ junction,  (b), and symmetric junction,
$\chi_l=-\chi_r$ (c). The positive and negative lobes of the
order parameter are chosen along $a$ and $b$ respectively (d).  }
\label{fig1}
\end{figure}
\begin{figure}[h]
\epsfysize 3cm \epsfbox[-200 230 300 570]{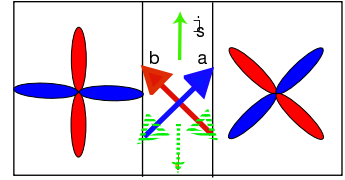}
\caption{Andreev levels and spontaneous currents in a DND model.
The total superconducting current across the boundary, carried by
normal levels (a) and $\protect\pi$-levels (b), is zero in
equilibrium, while the spontaneous current $j_s$ parallel to the
boundary is finite. The degeneracy of the ground state is
illustrated by the freedom of choice of its direction.}
\label{Fig2}
\end{figure}


\begin{eqnarray}
I(\phi )=\frac{ek_{F}}{\pi}\sum_{n}\int_{-\pi /2}^{\pi /2}d\theta
\cos \theta \ n_{F}[ E_{n}(\phi ,\theta )] \frac{dE_{n}(\phi
,\theta )}{d\phi }. \nonumber
\end{eqnarray}
Here, the $x$-axis is chosen normal to the grain boundary (see
Fig.\ \ref{fig1}), $n_{F}(E)$ is the Fermi distribution function,
and $E_{n}(\phi ,\theta )$ is the energy of the $n$-th Andreev
level for an electron with incidence angle $\theta $. The
direction-dependent intrinsic phase of $\Delta $ in $d$-wave
superconductors leads to qualitatively new features of
surface/interface states in $d$-wave superconductors (for current
reviews see \cite {KashiwayaTanakaReview,LofwanderReview}). They
stem from the coexistence of two types of Andreev levels, which
are formed by Andreev reflections from the order parameter with
the same (regular levels) or different ($\pi $-levels) intrinsic
phases (see Fig.\ \ref{fig1}). In case of a $d$-wave
superconductor boundary with an insulator, the latter include so
called midgap states (MGS), or zero energy states
(ZES)\cite{Hu94}, exactly at the Fermi level (since for $\phi =
\pi$ the condition (\ref{BSquant}) is always satisfied by $E=0$).
The two sets of Andreev levels carry Josephson currents in
opposite directions. In clean SND or DND junctions, as well as in
short SD or DD point contacts, their contributions behave
otherwise similarly. As a result, the current-phase dependence
can become $\pi $-periodic \cite{Yip94,Tanaka,Zagoskin97}. What
is more interesting, the equilibrium phase \ difference across
such a junction is neither 0, nor $\pi $, but is degenerate, $\pm
\phi _{0}$, which reflects the $\mathcal{T}$-breaking ground
state of the system \cite{Huck}. The value of $\phi _{0}$ at $T=0$
depends only on the orientation of the $d$-wave order parameter
with respect to the boundary \cite{Zagoskin97}. The occurrence of
spontaneous currents in this picture is natural \cite{Huck} (Fig.
\ref{Fig2}). In equilibrium, the Josephson current between the
left and right superconductors (in the $x$-direction) is zero,
which is possible only if the contributions of regular and $\pi
$-levels cancel each other. But the $y$-components \ of these
contributions add up, yielding a spontaneous current
\textit{along} the boundary. Evidently, there are two equilibrium
states (with a spontaneous current flowing up or down).

This picture becomes more complicated in the presence of finite
boundary transparency $D$, which breaks the symmetry between zero
and $\pi$-states. The latter are not split by specular tunneling,
and as a result, the amplitude of supercurrent carried by them
scales as $\sqrt{D}$ due to resonant transmission
\cite{RiedelBagwell} (instead of $D$ for regular levels). This led
to the prediction of temperature-driven transition to
$\pi$-junction \cite{Barash} in a DD junction. Recent
measurements of $I(\phi )$ in YBCO grain boundary junctions
\cite{Ilichev01} generally confirmed this prediction, but showed
that the transition does not necessarily occur between
equilibrium values $\phi _{0}=0$ and $\phi _{0}=\pi $.

Notice that in all of the above arguments we did not assume any
subdominant order parameter. In other words, the
$\mathcal{T}$-breaking and generation of the spontaneous current
happens without any subdominant pairing potential. In general, a
subdominant order parameter will occur at the boundary, if there
exist interactions in the corresponding channel. In DD-junctions,
the source of the subdominant order parameter is the proximity to
a differently oriented superconductor and therefore it will exist
at all temperatures even above the subdominant critical
temperature $T_{c2}$. It is interesting that the presence of a
subdominant interaction channel will actually suppress the
spontaneous current \cite{AOZ01}. We shall discuss this
counterintuitive behavior in section \ref{Spontaneous}.

The role of the subdominant order parameter in the spontaneous
current generation at surfaces is completely different. Near a
boundary [e.g.\ the (110)-surface] the dominant $d_{x^{2}-y^{2}}$
order parameter is suppressed \cite{Iguchi} and therefore  a
subdominant order parameter ($d_{xy}$ or $s$) can appear. It will
be formed below the smaller critical temperature $T_{c2}$ if
there exists nonzero pairing interaction in the corresponding
channel \cite{LLIX}. The combination of the two order parameters
with complex coefficients breaks the $\mathcal{T}$-symmetry
\cite{SigristRice92} and can lead to spontaneous surface currents
and magnetic fluxes. Usually $d_{x^{2}-y^{2}}\pm is$ or
$d_{x^{2}-y^{2}}\pm id_{xy}$ combinations are predicted. Recent
observations of zero bias peak splitting in surface tunneling
experiments \cite {Covington} and spontaneous fractional flux
(0.1-0.2 $\Phi _{0}$) near the ``green phase'' inclusions in YBCO
films \cite {TafuriKirtley} agree with this picture.

The simple description based on Andreev levels presented earlier
is best suited for quasi-1D problems in SNS (or DND) structures.
It is possible to generalize it to deal with at least some of the
above mentioned complications \cite{BZ99,DDqubit} by introducing
``Andreev tubes'' of width $\sim \lambda _{F}$, following the
quasiclassical trajectories. Nevertheless, in order to obtain
quantitative results, it is better to use an approach based
directly on the method of quasiclassical superconducting Green's
functions (Eilenberger equations \cite{Eilenberger69}). In this
paper we apply the formalism to the case of a planar $d$-wave
superconductor in contact with another superconductor or vacuum,
for an arbitrary transparency and roughness of the boundary.

The paper is organized as follows. In Section \ref{Spontaneous1},
we consider the generation of spontaneous currents by the
proximity effect at a uniform SD or DD interface. We obtain the
equilibrium phase, the spontaneous current distribution, and the
superscreening effect (in the latter case) in case of ideal,
rough, and partly reflective surface. In Section
\ref{Spontaneous}, the spontaneous current generated by the
subdominant order parameter at a boundary is considered. The
interplay of the two mechanisms (proximity effect and subdominant
pairing) is discussed for DD and SD junctions. The technical
details of the formalism are given in the appendices.

\vspace{-2mm}
\section{Spontaneous current generated by proximity effect}
\label{Spontaneous1}

\vspace{-2mm}
\subsection{Ideal junctions}

Let us consider a planar $d$-wave superconductor with a straight
grain boundary along the $y$-axis in its $ab$-plane (cf.\ Fig.\
\ref{fig1}). The order parameter $\Delta ({\bf v}_{F},{\bf r})$
is self-consistently determined by the interaction potential
$V({\bf v}_{F},{\bf v}_{F}^{\prime })$ [see Eq.\ (\ref{GapEq})].
In a pure $d_{x^2-y^2}$ case we assume the interaction potential
to have the form

\begin{equation}
V({\bf v}_{F},{\bf v}_{F}^{\prime })=V_{d}\cos 2(\theta-\chi) \cos
2(\theta'-\chi),  \label{Eq2.1}
\end{equation}
where the angles $\theta ,\theta'$ give the direction of ${\bf
v}_{F},{\bf v}_{F}^{\prime }$ in the $ab$-plane, and $\chi$ is the
angle between the crystallographic $a$-direction and the $x$-axis.
The dimensionless BCS constant of interaction is $\lambda
_{d}=V_{d}N(0)/2.$ We can also consider a boundary between a
$d$-wave superconductor and an $s$-wave film in which case, for
the $s$-wave superconductor, we have

\begin{equation}
V({\bf v}_{F},{\bf v}_{F}^{\prime })=V_{s}, \quad \lambda
_{s}=V_{s}N(0). \label{Eq2.2}
\end{equation}

This problem is essentially 1-dimensional, with $\Delta ({\bf
v}_{F},x) \rightarrow \Delta _{r,l}({\bf v}_{F})$ as $x\rightarrow
\pm \infty $, where the subscripts $l$ and $r$ represent left and
right of the boundary respectively. The method we choose to solve
this problem is the standard quasiclassical method which is
described in detail in appendix \ref{QCEEq.}. The self-consistent
solution can be obtained only numerically (see appendix
\ref{Num}). However, we start from the analytical solution in the
simplest approximation, when we assume that the order parameter
takes its bulk values at all $x$, there is no subdominant order
parameter, and the grain boundary is ideal (transparent and
specular, see appendix \ref{Anl}).

We denote by $j_{J}$ [$\equiv j_x(x{=}0)$] the Josephson current
flowing from the left to the right superconductor, and by $j_{S}$
[$\equiv j_y(x{=}0)$] the surface current flowing along the
interface at the boundary. All the current distributions are
expressed in units of $j_c$ [defined in (\ref{jc})] which is of
the order of the bulk critical current density. These currents
are expressed by Eqs. (\ref{Eq2.9}) and (\ref{Eq2.10}) which are
valid (within the applicability of the model) for arbitrary
symmetry of the order parameters $\Delta _{l,r}$.  For a DD
interface, the functions $\Delta _{l,r}({\bf v}_{F})$ in
(\ref{Eq2.9}) and (\ref{Eq2.10}) are
\begin{equation}
\Delta _{l,r}=\Delta _{0}(T)\cos 2(\theta -\chi _{l,r}),
\label{dd}
\end{equation}
where $\Delta _{0}(T)$ is the maximum gap, as introduced in
appendix \ref{OP}. Note that in this section the pairing
potential is assumed to be nonzero only in a single orbital
channel on either side of the boundary. Therefore, while the
anomalous Green's functions $(f,f^{\dagger })$ which support {\em
different} orbital symmetries are induced across the boundary,
they do not translate into a subdominant order parameter $\Delta
^{\prime }.$

\begin{figure}[h]
\epsfysize 4cm \epsfbox[20 260 300 510]{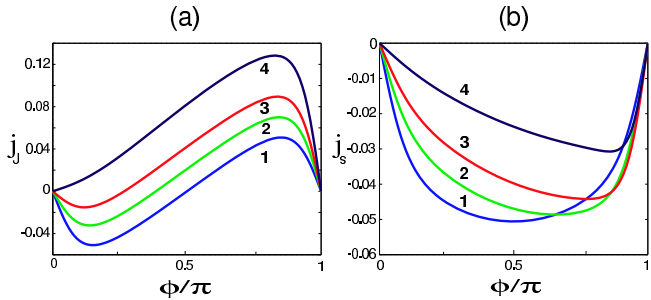}
\caption{Josephson current (a) and spontaneous current (b) versus
the phase difference in a clean DD grain boundary junction
calculated in non-self-consistent approximation. Current
densities are in units of $j_c$ [cf.\ Eq.\ (\ref{jc})] and
$T=0.1T_{c}$. The mismatch angles are $\chi_l=0$ and $\chi_r =
45^\circ$ (1),\ $40^\circ$ (2),\ $34^\circ$ (3),\ $22.5^\circ$
(4).} \label{fig3}
\end{figure}

The results of the calculations for a DD junction are displayed in
Fig.\ \ref{fig3} for different mismatch angles between the
crystalline axes across the grain boundary and at temperature
$T=0.1T_{c}$ (assuming the same transition temperature on both
sides). In all these figures, the left superconductor is assumed
to be aligned with the boundary while the orientation of the right
superconductor varies. The Josephson current-phase relation
(Fig.\ \ref{fig3}a) demonstrates a continuous transition from a
$\pi $-periodic (sawtooth-like) line-shape at $\delta \chi =
45^\circ$ to a 2$\pi$-periodic one for small $\delta \chi$, as
expected in the case of a clean DND junction \cite{Zagoskin97}.
The phase dependence of the surface current (Fig.\ \ref{fig3}b)
is also in qualitative agreement with earlier results for SND and
DND junctions \cite{Huck}.

It is important to make a cautionary remark here. In order to
have the Josephson effect, there must be a weak link between the
two superconductors. In other words, the superconducting phase
should change over a short distance at the boundary (otherwise
one should speak about phase gradient and not phase difference).
The weak link in a clean DND junction is provided by the normal
layer, while in an ideal DD junction, it is formed due to the
suppression of the order parameter near the boundary (at finite
$\delta \chi$), which follows from self-consistent treatment. At
$\delta \chi =0$, such a ``junction" (of infinite width) is
simply a {\em bulk} superconductor and not a weak link, therefore
Josephson physics does not apply. Nevertheless, the
non-self-consistent approximation is applicable, in the limit
$\delta \chi \to 0$, to the case of a narrow contact (microbridge)
between the two sides, in a way similar to conventional
superconductors \cite{KO}.

\begin{figure}[h]
\epsfysize 5cm \epsfbox[-20 180 400 600]{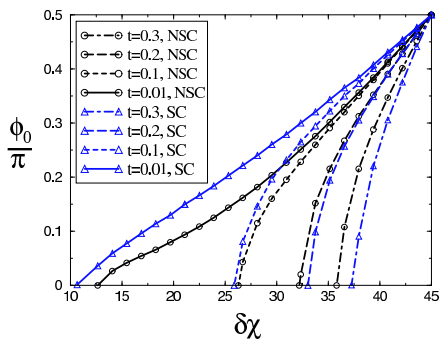}
\caption{Equilibrium phase difference in a clean DD grain
boundary junction as a function of $\protect\delta \protect\chi
\equiv \chi_r- \chi_l$ (keeping $\chi_l=0)$, at different values
of $t\equiv T/T_{c}.$ The circles and triangles correspond to
non-self-consistent (NSC) and self-consistent (SC) calculations
respectively. For nonzero $\protect\phi _{0},$ the ground state
is twice degenerate ($\phi=\pm \phi_0$).} \label{Fig4}
\end{figure}

\begin{figure}[h]
\epsfysize 5cm \epsfbox[-20 180 400 600]{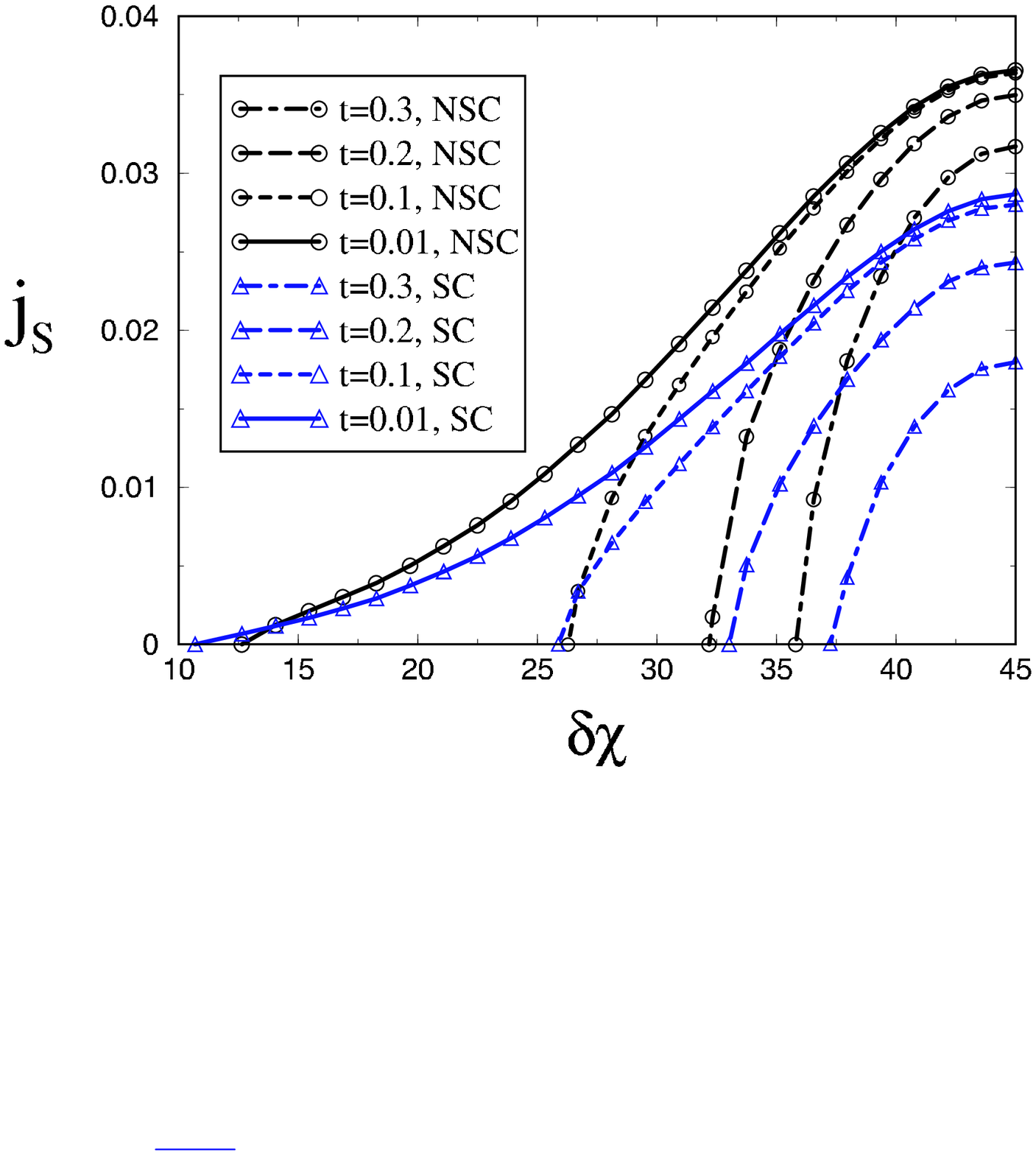}
\caption{Spontaneous current in the junction of Fig.\ \ref{Fig4}.}
\label{Fig5}
\end{figure}

The equilibrium phase difference across the junction, at which
$j_{J}(\phi )=0$ and $dj_{J}(\phi )/d\phi >0,$ takes any value
between $0$ and $\pi /2$, and is degenerate, $\phi =\pm
\phi_{0},$ unless $ \phi_0 =0.$ The $\mathcal{T}$-symmetry is
therefore broken. The region of $\mathcal{T}$-breaking states (as
a function of temperature and mismatch angle) is shown in Fig.
\ref{Fig4}. In that figure we also present the self-consistent
numerical result for comparison. The method adopted for our
numerical calculations is described in detail in appendix
\ref{Num}. Only in the asymmetric $\delta \chi =45^\circ$ junction
does the degeneracy (at $\phi= \pm \pi /2$) survive at all
temperatures, due to its special symmetry which leads to complete
suppression of all odd harmonics of $I(\phi)$; generally, $\phi
_{0}\rightarrow 0$ at some temperature that depends on the
orientation. The equilibrium value of the spontaneous current is
nonzero in a certain region of angles and temperatures (Fig.
\ref{Fig5}), which is largest in the case of asymmetric $\delta
\chi =45^\circ$ junction.

\begin{figure}[h]
\epsfysize 5cm \epsfbox[30 250 400 570]{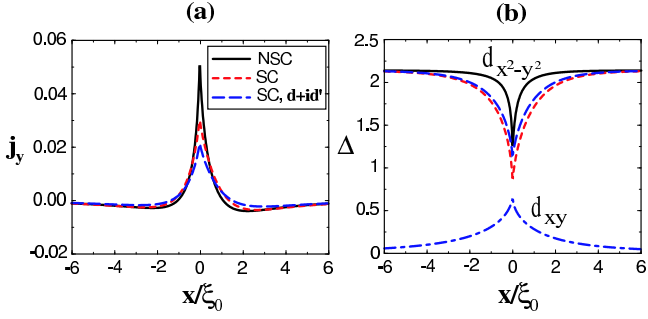}
\caption{Equilibrium spontaneous current distribution (a) and
order parameter (b) for a ($0^\circ$-$45^\circ$) grain boundary
junction at $t=0.1$. Solid and dashed lines represent the results
of non-self-consistent and self-consistent calculations
respectively. Long-dashed and dot-dashed lines represent the case
with interaction in the subdominant $d_{xy}$ channel
($t_{c2}\equiv T_{c2}/T_{c} =0.05$). } \label{fig6}
\end{figure}

We can also calculate the function $j_{y}({\bf r})$ at all $x$
using our analytical formalism [cf.\ Eq.\ (\ref{j(x)})]. This
function is plotted in Fig.\ \ref{fig6}a. In this figure we also
plot the same graph calculated using the self-consistent
numerical method described in appendix \ref{Num}, with and
without the subdominant order parameter (subdominant pairing is
discussed in appendix \ref{OP}). The curves are qualitatively
similar although there is a small quantitative difference. The
same is true for the dominant order parameter, but not for the
subdominant order parameter which appears only when the
interaction in the corresponding channel exists (cf.\ Fig.\
\ref{fig6}b, $\Delta$ in this figure and all subsequent figures
is normalized to $T_c$). Notice that the subdominant order
parameter exists near the boundary despite $T>T_{c2}$. The reason
is that the appearance of the subdominant order parameter is
merely the result of the proximity effect, i.e.,  the extension
of the order parameter from one region to the other, which
happens at all temperatures.

In all of the graphs in Fig.\ \ref{fig6}a we see a remarkable
feature: the current along the boundary is sharply peaked in a
layer of order $\xi _{0}$ around it, but is accompanied by
counterflows spread over about $10\xi _{0}$ on either side.
Within the numerical accuracy, the total current in $y$-direction
is \textit{zero,} independently on either side of the junction.
Since in high-$ T_{c}$ superconductors $10\xi _{0}\ll \lambda
_{L},\lambda _{J}$ (London and Josephson penetration depth
respectively), the phenomenon can be called
\textit{superscreening}\cite{AOZ01}. Note that we so far did not
take into account the Meissner screening -- and as it turns out
don't need to; the magnetic field of spontaneous currents is
cancelled on a scale less than $\lambda _{L}.$ This effect may be
responsible for difficulties with observing spontaneous currents
at surfaces and interfaces of $d$-wave superconductors
\cite{VH01}, although as we will see, the effect is suppressed by
roughness or reflectivity of the boundary.

\begin{figure}[h]
\epsfysize 5cm \epsfbox[40 250 400 560]{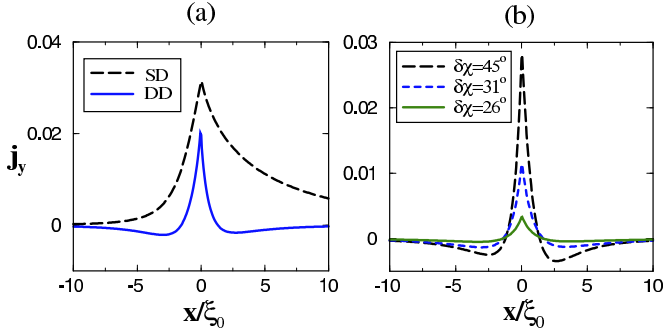} \caption{(a)
Spatial distribution of spontaneous current density in SD and
asymmetric DD clean junctions at $t=t_{c2}=0.05$ and
$t_{cs}=0.1$. In both cases $\chi_l=45^\circ$ and in the DD case
$\chi_r=0$. (b) Spontaneous current near DD junctions with
different misorientation angles $\chi_l=0$ and $\chi_r=\delta
\chi$ at $t=0.1$. Calculations in this figure and in all
subsequent figures are done self-consistently.} \label{fig7}
\end{figure}

\begin{figure}[h]
\epsfysize 4cm \epsfbox[-50 270 300 600]{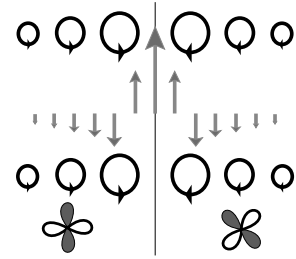}
\caption{``Molecular currents'': the supercurrent distribution in
a DD junction as the superposition of local currents produced by
the angular momenta of Cooper pairs in a chiral $ d+e^{\pm
i\protect\phi _{0}}d^{\prime }$ state.} \label{fig8}
\end{figure}

In the case $\delta \chi =45^\circ$, the superscreening can be
obtained analytically from Eq.\ (\ref{j(x)}); the integral
\begin{eqnarray}
\int_{0}^{\pm \infty }&& dx \ j_{y}(x)\propto \\
&& T\sum\limits_{\omega
>0} \left\langle \frac{\Delta _{l}\Delta _{r}\sin \theta \
{\rm sign}  (\cos \theta )\sin \phi }{\Omega _{l}\Omega _{r}
+\omega ^{2}+\Delta _{l} \Delta _{r} \cos \phi }\cdot \frac{\left|
v_{F}\cos \theta \right| }{\Omega _{r}} \right\rangle _{\theta }
\nonumber
\end{eqnarray}
is exactly zero after angular averaging. However, self-consistent
numerical calculations show (within the numerical accuracy) the
same behavior at all orientations (Fig.\ \ref{fig7}b). To better
understand the situation, let us recall that in a system with
local magnetic moment density ${\bf m(r)}$ the ``molecular
currents'' flow with density ${\bf j(r)=}c{\bf \nabla \times
m(r)}$ (Fig.\ \ref {fig8}). In a superconductor with order
parameter $d_{x^{2}-y^{2}}+e^{i \phi _{0}}d_{xy}$, the local
orbital/magnetic moment density is
\begin{eqnarray*}
{\bf l(r)} &\propto& {\bf m(r)\propto} \ \widehat{{\bf z}}%
\int_{0}^{2\pi }\frac{d\theta }{2\pi }\left( \Delta _{1}(x)\cos
2\theta +\Delta _{2}(x)e^{-i\phi _{0}}\sin 2\theta \right) \\
&& \times \frac{1}{i}\frac{\partial }{\partial \theta }\left(
\Delta _{1}(x)\cos 2\theta +\Delta _{2}(x)e^{i\phi
_{0}}\sin 2\theta \right) \\
&=&2\Delta _{1}(x)\Delta _{2}(x)\widehat{{\bf z}}\sin \phi _{0}.
\end{eqnarray*}
The contribution to the spontaneous current is thus ${\bf
j(r)\propto \nabla \times l(r)\parallel \hat{y}.}$ Note that the
same expression is obtained from the Ginzburg--Landau
equations\cite{BSL97}: ${\bf j}\propto \nabla \times {\bf
\hat{z}}\ {\rm Im} [d_{1}\left( {\bf r}\right) d_{2}^{\ast }\left(
{\bf r}\right)].$ The total current in the $y$-direction due to
this mechanism is $I_{tot}\propto $ $\int_{\Omega }d{\bf S\cdot
\nabla \times m\propto }\oint_{\partial \Omega }d{\bf s\cdot m},$
where $\Omega $ is a cross-section in the $xz$-plane (assuming the
superconducting film has finite thickness $h$), and $d{\bf s}$ is
the linear element of its boundary $\partial \Omega .$ The latter
integral is obviously zero, because either ${\bf m \parallel
\hat{z} \perp }d{\bf s}$ (when $z=\pm h/2$), or ${\bf m}=0$ (at
$x=\pm \infty $ , where the contour $\partial \Omega $ is closed).
Moreover, the ``molecular currents'' picture, shown in Fig.
\ref{fig8}, reproduces to the distinct peak-and-counterflow
current distribution.

A finite magnetic moment in the bulk is essential for the effect.
The corresponding chiral state is created by the proximity effect,
when the two $d$-wave order parameters mixed near the boundary
are spatially rotated by a nontrivial angle $\delta \chi $ and
have a nontrivial phase shift $\pm \phi _{0}\neq 0,\pi$. Although
the argument given here is based on having a mixed order parameter
near the boundary region, in fact the presence of the second
order parameter is not necessary. The reason is that the
subdominant symmetry already exists in the Green's functions
$g_\omega$ and $f_\omega$, even if the interaction potential does
not support pairing in that channel. Since the current density is
related to these functions rather than to the pairing potential
[cf.\ Eq.\ (\ref{EqA6})], the molecular current is basically
formed by inherently mixed symmetries of these functions.

The above arguments are certainly not true when the order
parameter symmetry is $d+is$, as in SD junctions (Fig.
\ref{fig7}a). In this case the molecular current is identically
zero and as a result no countercurrent exists. (Of course, since
the Meissner currents must be taken into account in this case,
the results are valid only if the system size is much less than
the London penetration depth.)

\begin{figure}[h]
\epsfysize 4.2cm \epsfbox[-30 210 400 550]{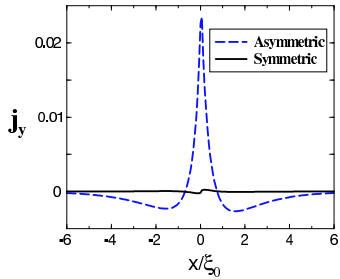}
\caption{Comparison between the size of the spontaneous current in
asymmetric and symmetric $\delta \chi = 45^\circ$ DD junctions.
Both calculations are done at $t=0.5$.} \label{fig9}
\end{figure}

An interesting case is presented by a symmetric
$45^\circ$-junction ($\chi_l=-22.5^\circ$ and
$\chi_r=22.5^\circ$). Although the ground state is degenerate in
this case \cite{Ilichev01}, the spontaneous current is
practically absent (Fig.\ \ref{fig9}). This is easy to see from
geometric considerations (Fig.\ \ref{fig1}c). Consider the total
supercurrent across the boundary as a sum of contributions from
quasiclassical trajectories. There are two groups, analogous to
Andreev regular and $\pi$-levels discussed in the introduction;
the corresponding directions are within the shaded and white
sectors in Fig.\ \ref{fig1}c respectively. In equilibrium, their
contributions to the current normal to the boundary must cancel
each other. What happens to the tangential component of the
equilibrium (spontaneous) current, depends on the orientations of
the order parameters with respect to the boundary and each other,
and the equilibrium phase difference (also a function of
temperature). It is obvious from the picture that for any given
mismatch angle $\delta \chi $, in a \emph{symmetric} junction,
the tangential components of the current cancel for regular and $
\pi $-directions \emph{separately}. This is why symmetric
junctions are very ``quiet'' (using the term of \cite{IOFFE}):
they can violate $\mathcal{T}$-symmetry without producing local
magnetic fields, which could couple to some external degrees of
freedom\cite{HOneill}. This can be both an advantage and a
disadvantage from the point of view of using such a system as a
solid-state qubit \cite{IOFFE,DDqubit}.

\vspace{-2mm}
\subsection{Reflective junctions}

We also study the properties of the spontaneous current in the
presence of a non-ideal boundary, i.e., with nonunity
transparency and/or surface roughness. The details of our method
of numerical calculation are given in appendix \ref{Num}. The
transparency and roughness of the boundary are parameterized by
$0 \le D_0 \le 1$ and $0 \le \rho < \infty$ respectively (see
appendix \ref{Num} for definitions). In particular, in the case
of a clean ideal junction, $D_0=1$ and $\rho =0$.

\begin{figure}[h]
\epsfysize 5cm \epsfbox[40 260 400 580]{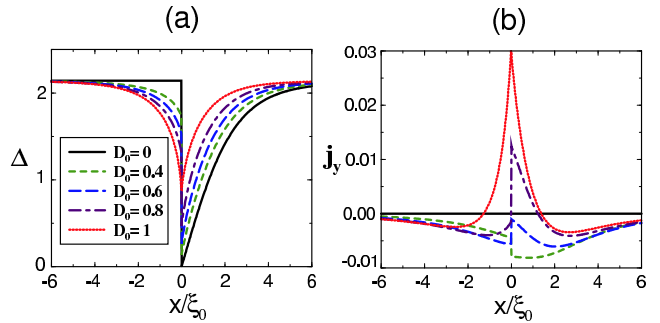} \caption{Order
parameter (a) and spontaneous current distribution (b) in
($0^\circ$-- $45^\circ$) asymmetric grain boundary junctions with
different transparencies $D_0$ at $t=0.1$.} \label{fig10}
\end{figure}

\begin{figure}[h]
\epsfysize 4.5cm \epsfbox[30 250 400 540]{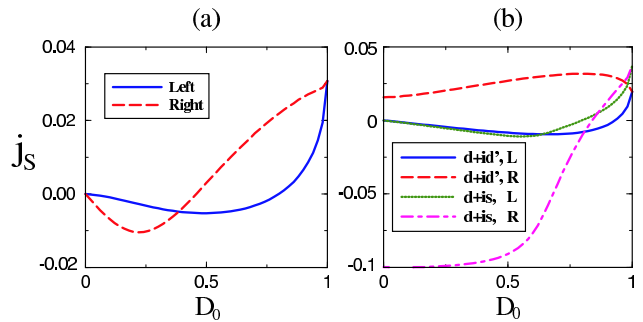}
\caption{Influence of finite barrier transparency on spontaneous
current amplitude in an asymmetric ($0^\circ$-$45^\circ$) DD
junction without (a) and with (b) subdominant order parameter.
Here ``L"= {\em left} and ``R"= {\em right} of the boundary and
the calculations are done at $t=0.05$ and $t_{c2}=0.1$ (for (b)).
In the case of (b), one can see that as $D_0\rightarrow 0$, the
spontaneous currents on the right hand side of the boundary
survive.} \label{fig11}
\end{figure}

In Fig.\ \ref{fig10}a we see the self-consistent pairing potential
near the interface for several values of $D_0$. Notice that for
the case of $D_0=0$ (i.e.,  totally reflecting surface), the order
parameter is constant on the left while it vanishes at the
junction on the right side of the boundary. The latter is because
the quasiparticles on the right see opposite sign of the order
parameter after reflection from the boundary \cite{Matsumoto}. As
$D_0 \rightarrow 1$, the order parameter becomes continuous at
the boundary. The spontaneous current (Fig.\ \ref{fig10}b) also
evolves from zero to its maximum value as $D_0$ changes from 0 to
1. Notice that at all values of $D_0$ except for 0 and 1, the
magnitudes of the spontaneous current on the left and right hand
sides of the boundary are different and exact superscreening
happens only in the ideal boundary case ($D_0=1$). Fig.\
\ref{fig11}a demonstrates the continuous evolution of the
spontaneous current from zero to its maximum as a function of
$D_0$. This picture is influenced by the presence of a subdominant
interaction in Fig.\ \ref{fig11}b. Especially, on the right hand
side of the junction (with $\chi_r=45^\circ$ orientation), the
spontaneous current does not vanish even at $D_0=0$. We will come
back to this point in the next section.

\vspace{-2mm}
\subsection{Rough junctions}

\begin{figure}[h]
\epsfysize 4.5cm \epsfbox[30 290 400 580]{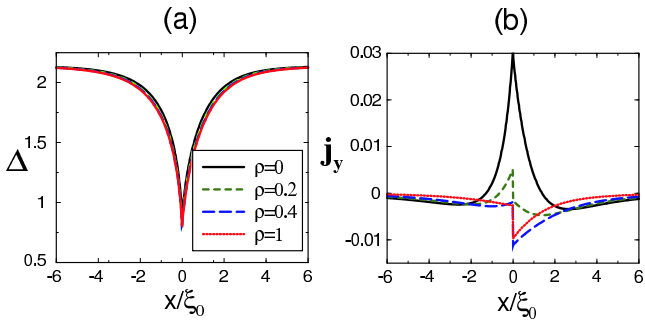}
\caption{Order parameter (a) and spontaneous current distribution
(b) in ($0^\circ$-$45^\circ$) asymmetric grain boundary junction
with different values of the surface roughness $\rho$ at $t=0.1$.}
\label{fig12}
\end{figure}

The effect of surface roughness is very different from that of
finite transparency. First notice from Fig.\ \ref{fig12} that
surface roughness has little effect on the order parameter
distribution (Fig.\ \ref{fig12}a), while it significantly changes
the spontaneous current (Fig.\ \ref{fig12}b). This may seem
strange, until we recall that supercurrents (including spontaneous
currents) are generated not by the order parameter $\Delta $, but
by the Green's functions, $f_\omega$ and $g_\omega$ [related by
the normalization condition (\ref{EqA4})]. The latter $\emph{is}$
directly affected by the boundary roughness, while the former
isn't. Again, we notice that exact superscreening is absent at
finite roughness.

\begin{figure}[h]
\epsfysize 4.5cm \epsfbox[20 240 400 530]{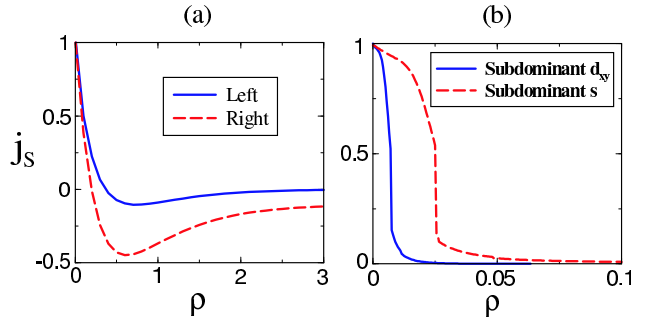}
\caption{Influence of surface/interface roughness on the
magnitude of spontaneous current in a junction (a) and near a
surface (b). Note that in the latter case spontaneous currents
are suppressed at much smaller values of $\protect\rho$. All
calculations are done at $t=0.1$} \label{fig13}
\end{figure}

Another interesting effect is presented in Fig.\ \ref{fig13}a: as
the surface becomes rougher and rougher ($\rho \rightarrow
\infty$), the spontaneous current on the left side (with
$\chi_l=0$) vanishes while on the right side ($\chi_r=45^\circ$)
it saturates to a finite value. This non-trivial behavior is
directly related to the anomalous proximity effect between a
$d$-wave superconductor and a disordered region studied by
Golubov and Kupriyanov \cite{golubov}. When $\rho$ is large, the
contributions from different quasiclassical trajectories are
mixed. Therefore the angle-dependent components of $f$, which
``leak'' through the boundary to the other side, are suppressed,
but in the limit of infinite roughness the $s$-wave contribution
still survives. The latter is generated by scattering from the
rough boundary even in case of only dominant $d$-wave pairing on
either side, and can be estimated as $f_{s,\rm eff}=\int_{-\pi
/2}^{\pi /2}(d\theta/\pi)\cos \theta f(\theta ,0^{\pm })$ [the
upper (lower) sign corresponds to leakage from the right (left)
to the left (right) side of the junction]. In an asymmetric
($0^\circ$-$45^\circ$) DD junction we thus expect exactly zero
induced $f$ on the left due to symmetry, while on the right a
finite $s$-wave component should appear with amplitude $2/(3\pi
)\approx 0.2$ relative to the $d$-wave anomalous Green's function
across the boundary. The spontaneous currents in this case can
flow only on the right of the boundary, where the chiral
combination $d \pm is$ is thus formed, in agreement with our
numerical results. This might also explain why superscreening is
absent in case of rough surfaces: $d+is$ symmetry does not supprot
molecular moments. Non-ideal asymmetric grain boundaries between
$d$-wave superconductors can thus be better candidates for the
search of spontaneous currents, since the superscreening is
suppressed, while the supercurrent amplitudes are still
detectable.

\vspace{-2mm}
\section{Spontaneous current generated by subdominant order parameter}
\label{Spontaneous}

We already saw the effect of the presence of a subdominant order
parameter on the spontaneous current in DD grain boundary
junctions (Fig.\ \ref{fig6}). In particular in Fig.\ \ref{fig6}b
we notice that the subdominant component exists even at
temperatures above the subdominant critical temperature, $T_{c2}$.
The subdominant order parameter in Fig.\ \ref{fig6}b is therefore
completely induced by the proximity to a differently oriented
superconductor on the other side of the junction. On the other
hand, in Fig.\ \ref{fig11}b we realized that the spontaneous
current exists even when $D_0 \rightarrow 0$, i.e.,  when the
proximity effect is completely absent. This happens only when $T
< T_{c2}$ and therefore the mechanism is completely different
from the one described above.

\begin{figure}[h]
\epsfysize 5cm \epsfbox[30 230 300 550]{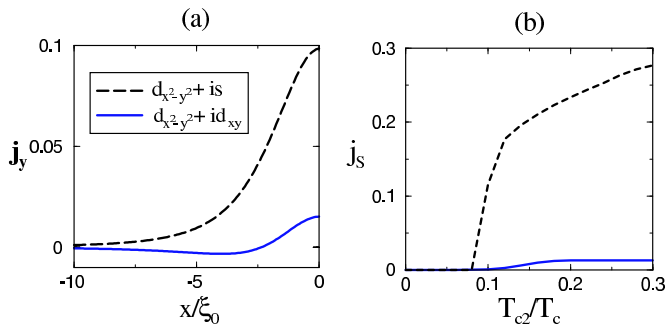} \caption{(a)
Spontaneous current profile near the (110)-surface of a $d$-wave
superconductor with $d_{xy}$- and $s$-wave subdominant order
parameter. Here $t=0.05$ and $t_{c2}=0.1$ for both the $s$ and
$d_{xy}$ subdominant gaps. (b) Spontaneous current at the surface
as a function of the subdominant transition temperature at
$t=0.1$.} \label{fig14}
\end{figure}

In general, the bulk of a superconductor can support only one
symmetry of the order parameter even if the interaction potential
contains finite interaction in several different channels (unless
the coupling constants of the channels are very close to each
other \cite{herbut}). This is because the dominant order parameter
introduces a cutoff, in the BCS gap equation, that removes the
logarithmic divergence (at small $\omega$), responsible for the
exponential dependence of the gap on the interaction potential
(see appendix \ref{OP}). As we saw in Fig.\ \ref{fig10}a, at a
reflective surface of a $45^\circ$ oriented $d$-wave
superconductor, the dominant order parameter vanishes.
Suppression of the dominant order parameter gives the chance to
the next subdominant one to appear if the temperature is below
its corresponding critical temperature. Thus, the appearance of
subdominant order involves a spontaneous symmetry breaking and
therefore a second order phase transition. This is demonstrated
in Fig.\ \ref{fig14}b (we increase $T_{c2}$ instead of $T$ in
that figure, but the behavior is the same). The subdominant order
parameter always appears with a phase $\pi/2$ with respect to the
main one. With this phase difference, the mixed order parameter
($d_{x^{2}-y^{2}}\pm id_{xy}$ or $d_{x^{2}-y^{2}}\pm is$) is
fully gapped (with no nodes), which is energetically favorable
because of the extra condensation energy gained.

It is important to notice that the above phenomenon is not likely
to happen at the interface between superconductors because the
dominant order parameter is never completely suppressed (except
when $D=0$, cf.\ Fig.\ \ref{fig10}a). Even at rough interfaces,
the dominant order parameter at the boundary is only slightly less
than half of its bulk value (Fig.\ \ref{fig12}a).  The subdominant
order parameter in this case is predominantly induced by the
proximity effect and therefore follows the symmetry of the
neighboring superconductor. This may explain why the mixed
symmetry state was not observed in a recent experiment
\cite{NV01}.

Spontaneous currents at a surface of a $d$-wave superconductor
are presented in Fig.\ \ref{fig14}a. Qualitatively, they are the
same as at a DD or DS boundary, with the same superscreening
behavior in the case of $d+id'$ order parameter symmetry. Despite
the similarities in the current distribution, there are
fundamental differences between the two cases. First of all,
unlike at the interface, the appearance of spontaneous current
near the surface is a very fragile phenomenon. A small deviation
from the $45^\circ$ angle will suppress the effect significantly.
It is also very sensitive to surface roughness as compared to the
interface current; a small $\rho$ is enough to remove the
spontaneous current (Fig.\ \ref{fig13}b). Moreover, unless the
effects of the magnetic field generated by spontaneous currents
itself are taken into account, the presence of a subdominant gap
is necessary for the surface spontaneous current, but not for the
interface one. (The spontaneous symmetry breaking due to Doppler
shifts of midgap states \cite{Higashitani} occurs at temperatures
below $1/6 \Delta (\xi_0/\lambda_L)$
\cite{Honerkamp,BarashTT,Lofwander2}, that is approximately 1K
for high-$T_c$ compounds.)

\begin{figure}[h]
\epsfysize 5cm \epsfbox[40 250 300 560]{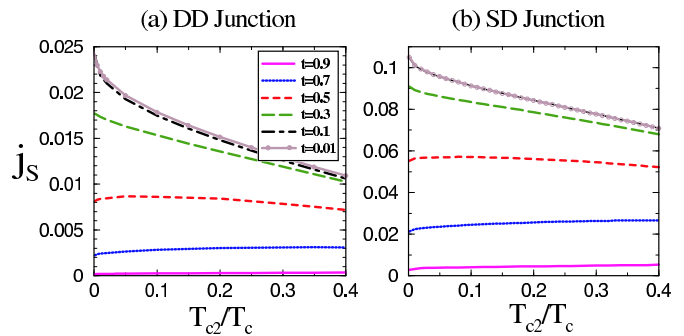}
\caption{Suppression of spontaneous currents in the presence of
subdominant order parameter as a function of subdominant
transition temperature. In both cases we have taken the same
$T_c$ on both side of the junction.} \label{Fig15}
\end{figure}

\begin{figure}[t]
\epsfysize 4.5cm \epsfbox[-450 -20 300 800]{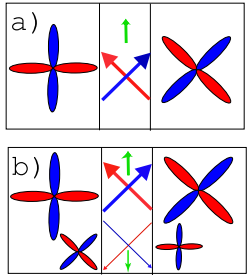} \caption{The
physical mechanism of spontaneous current suppression by
subdominant order parameter in DND model (see text).}
\label{Fig16}
\end{figure}

It is interesting to note that the presence of interaction in a
subdominant channel actually works {\em against} the appearance
of spontaneous current at an interface \cite{AOZ01}. This
behavior is displayed for both DD and SD junctions in Fig.\
\ref{Fig15}. The reason becomes clear from considering the
Andreev levels in the junction (which can be modeled on this
occasion by a DND junction, Fig.\ \ref{Fig16}). The induced
subdominant order parameter will be aligned with the dominant one
across the boundary. Therefore in addition to the
``dominant-dominant'' set of Andreev levels there will appear a
``subdominant-subdominant'' set, which obviously carries
supercurrents in the \textit{opposite }direction. Of course, a
subdominant pairing potential should also suppress the Josephson
current, which was indeed noted in \cite {TanKash}.

As we have mentioned above, the subdominant order parameter near
interfaces is induced across the boundary from the other side,
where it is dominant, and can be suppressed only simultaneously
with the latter. Therefore, it seems impossible to directly
observe the influence of the subdominant order on the spontaneous
currents by e.g.\ raising the temperature above $T_{c2}$ to
suppress it. The same, of course, holds true for suppression by
magnetic field.

\vspace{-2mm}
\section{Conclusions}

We have investigated the occurrence of spontaneous currents at
surfaces and interfaces of $d$-wave superconductors, using the
self-consistent quasiclassical Eilenberger equations. A
$\mathcal{T}$-breaking ground state is the only necessary
condition for their appearance, with no quantization condition
for the generated magnetic fluxes. Therefore, the effect is
sensitive to the properties of the system and allows the
existence of ``quiet'' $\mathcal{T}$-breaking states, i.e.,
states with fluxes much smaller than the flux quantum.

We have shown that the spontaneous current at the (110)-surface of
a $d$-wave superconductor is formed through a second order phase
transition to a mixed symmetry state ($d+id'$ or $d+is$), and is
very sensitive to the interaction in the subdominant channel,
temperature, and surface roughness and reflectivity. On the other
hand, at an interface between two differently oriented $d$-wave
superconductors, or between a $d$-wave and an $s$-wave
superconductor, the spontaneous current is generated as a result
of the proximity effect. It is generally robust, with less
sensitivity to the above effects. In particular, at very rough
surfaces, the spontaneous current survives on one side of the
junction, in agreement with the recently proposed anomalous
proximity effect at rough $d$-wave surfaces. We also show that
interaction in any subdominant channel suppresses the spontaneous
current at the interface, while its existence is necessary for
the $\mathcal{T}$-breaking at the surfaces.

The contribution to the spontaneous current from the local orbital
moment of the condensate leads to characteristic
``superscreening'' of the equilibrium current in DD junctions and
at surfaces with $d_{xy}$ subdominant order. Exact superscreening
is sensitive to boundary imperfections and disappears at finite
roughness and/or nonunity transparency of the junction.

\vspace{-2mm}
\section*{Acknowledgment}
We are grateful to I. Affleck, Ya. Fominov, A. Golubov, Yu. Gusev,
W. Hardy, E. Il'ichev, G. Rose, and A. Smirnov for enlightening
discussions, and A. Maassen van den Brink for carefully reading
this manuscript and many helpful remarks. M.A. would also like to
thank I. Herbut for stimulating conversations. A.Z. gratefully
appreciates valuable comments by T. L\"{o}fwander and J.
Hogan-O'Neill.

\appendix

\vspace{-2mm}
\section{Quasiclassical Eilenberger equations} \label{QCEEq.}

To describe the coherent current states in a superconducting
ballistic microstructure we use the Eilenberger equations for the
$\xi $-integrated Green's functions \cite{Eilenberger69}:
\begin{equation}
{\bf v}_{F}\cdot \frac{\partial }{\partial {\bf
r}}\widehat{G}_{\omega }({\bf v}_{F},{\bf r})+[\omega
\widehat{\tau }_{3}+\widehat{\Delta }({\bf v}_{F},{\bf
r}),\widehat{G}_{\omega }({\bf v}_{F},{\bf r})]=0,  \label{EqA1}
\end{equation}
where
\begin{equation}
\widehat{\Delta }=\left(
\begin{array}{cc}
0 & \Delta \\
\Delta ^{\dagger } & 0
\end{array}
\right), \quad \widehat{G}_{\omega }({\bf v}_{F},{\bf r})=\left(
\begin{array}{cc}
g_{\omega } & f_{\omega } \\
f_{\omega }^{\dagger } & -g_{\omega }
\end{array}
\right).
\end{equation}
$\Delta $ is the superconducting order parameter and
$\widehat{G}_{\omega }({\bf v}_{F},{\bf r})$ is the matrix Green's
function, which depends on the electron velocity on the Fermi
surface ${\bf v}_{F}$, the coordinate ${\bf r}$ and the Matsubara
frequency $\omega=(2n+1)\pi T$, with $n$ being an integer number
and $T$ the temperature. We also need to satisfy the
normalization condition
\begin{equation}
g_{\omega }=\sqrt{1-f_{\omega }f_{\omega }^{\dagger }}.  \label{EqA4}
\end{equation}
In general, $\Delta $ depends on the direction of ${\bf v} _{F}$
and is determined by the self-consistency equation

\begin{equation}
\Delta ({\bf v}_{F},{\bf r})=2\pi N(0)T\sum\limits_{\omega >0}
\left< V( {\bf v}_{F},{\bf v}_{F}^{\prime })f_{\omega }({\bf
v}_{F}^{\prime }, {\bf r}) \right>_{{\bf v}_{F}^{^{\prime }}}
\label{GapEq}
\end{equation}
where $V({\bf v}_{F},{\bf v}_{F}^{\prime })$ is the interaction
potential. Solution of matrix equation (\ref{EqA1}) together with
(\ref{GapEq}) determines the current density $ {\bf j(r)}$ in the
system

\begin{equation}
{\bf j(r)}=-4\pi ieN(0)T\sum\limits_{\omega >0} \left<{\bf
v}_{F}g_{\omega }({\bf v}_{F},{\bf r}) \right>_{{\bf v}_{F}}.
\label{EqA6}
\end{equation}
In two dimensions, $N(0)=m/ 2\pi$ is the 2D density of states and
$\left<... \right>=\int\limits_{0}^{2\pi } (d\theta / 2\pi )...$
is the averaging over directions of 2D vector ${\bf v }_{F}$.
Throughout this article we write current densities in units of
\begin{equation}
j_c \equiv 4\pi |e| v_F N(0) T_c \label{jc}
\end{equation}
which is of the order of the bulk critical current density.

Supposing $\Delta (-{\bf v}_{F})=\Delta ($ ${\bf v}_{F})$, which
is always the case for superconductors with singlet pairing, from
the equation of motion (\ref{EqA1}) and Eq.\ (\ref{EqA4}) we have
the following symmetry relations:
\begin{eqnarray*}
f^{\ast }(-\omega ) &=&f^{\dagger }(\omega ),\\
g^{\ast }(-\omega) &=& -g(\omega ),\\
f^{\ast }(\omega ,-{\bf v}_{F}) &=& f^{\dagger }(\omega ,{\bf v}
_{F}), \\
g^{\ast }(\omega ,-{\bf v}_{F})&=& g(\omega ,{\bf v}_{F}), \\
f(-\omega ,-{\bf v}_{F}) &=& f(\omega ,{\bf v}_{F}),\\
g(-\omega,-{\bf v}_{F})&=& -g(\omega ,{\bf v}_{F}), \\
\Delta ^{\dagger } &=&\Delta ^{\ast }.
\end{eqnarray*}

\vspace{-2mm}
\section{Non-self-consistent analytical solution}
\label{Anl}

We assume a boundary, at $x=0$, between two superconductors. To
find an analytical solution we neglect the self-consistency
equation (\ref{GapEq}), and write

\begin{equation}
\Delta ({\bf v}_{F},x)= \left\{
\begin{array}{ll}
\Delta _{l}({\bf v}_{F})\exp (i\phi /2), & x<0 \\
\Delta _{r}({\bf v}_{F})\exp (-i\phi /2), & x>0
\end{array} \right.
,  \label{Eq2.3}
\end{equation}
where $l$ ($r$) stands for left (right) of the boundary.
$\Delta_{l,r}$ can in general have $d$-wave, $s$-wave, or any
other symmetries. The Eilenberger equations for the Green's
function $\hat{G}_{\omega }$ are linear and therefore can be
easily solved separately for positive and negative $x$ to satisfy
\begin{equation}
\lim_{x\rightarrow \mp \infty }f_{\omega }=\frac{\Delta
_{l,r}}{\Omega _{l,r}},\qquad \lim_{x\rightarrow \mp \infty
}g_{\omega }=\frac{\omega }{\Omega _{l,r}},
\end{equation}
where $\Omega _{l,r}=\sqrt{\omega ^{2}+\left| \Delta
_{l,r}\right| ^{2}}$. This yields the following solutions. For
$x\leq 0$
\begin{eqnarray}
f(x,\theta ) &=&\frac{\Delta _{l}e^{i\phi /2}}{\Omega
_{l}}+\frac{e^{i\phi /2}}{\Delta _{l}}(\eta \Omega _{l}-\omega
)e^{2x\Omega _{l}/\left|
v_{x}\right| }C_{1}  \nonumber \\
f^{\dagger }(x,\theta ) &=&\frac{\Delta _{l}e^{-i\phi /2}}{\Omega
_{l}}+ \frac{e^{-i\phi /2}}{\Delta _{l}}(-\eta \Omega _{l}-\omega
)e^{2x\Omega
_{l}/\left| v_{x}\right| }C_{1}  \nonumber \\
g(x,\theta ) &=&\frac{\omega }{\Omega _{l}}+e^{2x\Omega
_{l}/\left| v_{x}\right| }C_{1},  \label{left}
\end{eqnarray}
for $x\geq 0$
\begin{eqnarray}
f(x,\theta ) &=&\frac{\Delta _{r}e^{-i\phi /2}}{\Omega
_{r}}+\frac{e^{-i\phi /2}}{\Delta _{r}}(-\eta \Omega _{r}-\omega
)e^{-2x\Omega _{r}/\left|
v_{x}\right| }C_{2}  \nonumber \\
f^{\dagger }(x,\theta ) &=&\frac{\Delta _{r}e^{i\phi /2}}{\Omega
_{r}}+ \frac{e^{i\phi /2}}{\Delta _{r}}(\eta \Omega _{r}-\omega
)e^{-2x\Omega
_{r}/\left| v_{x}\right| }C_{2}  \nonumber \\
g(x,\theta ) &=&\frac{\omega }{\Omega _{r}}+e^{-2x\Omega
_{r}/\left| v_{x}\right| }C_{2}.  \label{right}
\end{eqnarray}
Here $\eta \equiv {\rm sign}(v_{x})$. Imposing the continuity
condition for $\hat{G}_{\omega }$ at the boundary we find
\begin{eqnarray}
C_{1} &=&\frac{\Delta _{l}}{\Omega _{l}}\frac{\omega (\Delta
_{l}-\Delta _{r}\cos \phi )-i\eta \Delta _{r}\Omega _{l}\sin \phi
}{\Omega _{l}\Omega
_{r}+\omega ^{2}+\Delta _{l}\Delta _{r}\cos \phi}  \nonumber \\
C_{2} &=&\frac{\Delta _{r}}{\Omega _{r}}\frac{\omega (\Delta
_{r}-\Delta _{l}\cos \phi )-i\eta \Delta _{l}\Omega _{r}\sin \phi
}{\Omega _{l}\Omega _{r}+\omega ^{2}+\Delta _{l}\Delta _{r}\cos
\phi}.
\end{eqnarray}
Substituting into (\ref{left}) or (\ref{right}) at $x=0$ we find
the Green's functions $f$ and $g$ to be
\begin{eqnarray}
f(0)&=& \frac{\Delta _{l}(\Omega _{r}+ \eta \omega ) e^{i\phi/2}
+ \Delta _{r}(\Omega _{l}+ \eta \omega ) e^{-i\phi/2}}{\Omega
_{l}\Omega _{r}+\omega
^{2}+\Delta _{l}\Delta _{r}\cos \phi }, \\
g(0)&=& \frac{\omega (\Omega _{l}+\Omega _{r})-i\eta \Delta
_{l}\Delta _{r}\sin \phi }{\Omega _{l}\Omega _{r}+\omega
^{2}+\Delta _{l}\Delta _{r}\cos \phi }. \label{Eq2.8}
\end{eqnarray}

Using the expressions (\ref{GapEq}) and (\ref{Eq2.8}), we obtain
the current densities $j_{x}(0)\equiv j_{J}$ and $j_{y}(0)\equiv
j_{S}$
\begin{equation}
j_{J}= t \sum\limits_{\omega >0} \left\langle \frac{\Delta
_{l}\Delta _{r}\left| \cos \theta \right| }{\Omega _{l}\Omega
_{r}+\omega ^{2}+\Delta _{l}\Delta _{r}\cos \phi } \right\rangle
_{\theta }\sin \phi ,  \label{Eq2.9}
\end{equation}
\begin{equation}
j_{S}= t \sum\limits_{\omega >0} \left\langle \frac{\Delta
_{l}\Delta _{r}\sin \theta \ {\rm sign}(\cos \theta )}{\Omega
_{l}\Omega _{r}+\omega ^{2}+\Delta _{l}\Delta _{r}\cos \phi }
\right\rangle _{\theta }\sin \phi , \label{Eq2.10}
\end{equation}
where the current densities are in units of $j_c$ [defined in
(\ref{jc})] and $t \equiv T/T_c$. At all other points we can find
${\bf j}$ from

\begin{eqnarray}
{\bf j}(x) = t \sum\limits_{\omega >0} \left\langle
\frac{\widehat{v}_F \ \eta \Delta _{l}\Delta _{r} \sin \phi \ e^{
-{2\left| x\right| \Omega _{r}/ \left| v_{F}\cos \theta \right|
}} }{\Omega _{l}\Omega _{r} +\omega ^{2}+\Delta _{l} \Delta _{r}
\cos \phi } \right\rangle _{\theta } \label{j(x)}
\end{eqnarray}
where $\widehat{v}_F$ is a unit vector in the direction of ${\bf
v}_F$.

\vspace{-2mm}
\section{Self-consistent numerical solution}
\label{Num}

The general approach we use is based on transformation of the set
of coupled (formally) linear Eilenberger equations (\ref{EqA1})
for the functions $f,f^{\dagger },g$ to two nonlinear, but
separate, equations which are numerically stable (Schopohl-Maki
transformation \cite{Schopohl}). To this end we express the
components of $\hat{G}_{\omega }$ matrix as
\begin{equation}
g={\frac{1-ab}{1+ab}}\ ,\quad f={\frac{2a}{1+ab}}\ ,\quad
f^{\dagger }=\frac{2b}{1+ab}.  \label{EqC1}
\end{equation}
Now $a$ and $b$ satisfy two independent nonlinear equations
\begin{eqnarray}
{\bf v}_{F}\cdot \nabla a &=&\Delta -\Delta ^{\ast
}a^{2} - 2\omega a\nonumber \\
-{\bf v}_{F}\cdot \nabla b &=&\Delta ^{\ast }-\Delta b^{2}
-2\omega b . \label{EqC2}
\end{eqnarray}
Notice from these equations that $a(-{\bf v}_F)= b^*({\bf v}_F)$
and $b(-{\bf v}_F)= a^*({\bf v}_F)$. We use the solutions for a
homogeneous system,
\begin{equation}
a={\frac{\Delta }{\omega +\Omega }}\ ,\qquad b={\frac{\Delta
^{\ast }}{\omega +\Omega }}  \label{EqC3}
\end{equation}
where $\Omega =\sqrt{\omega ^{2}+|\Delta |^{2}}$, as asymptotic
conditions at $x\rightarrow \pm \infty $. For positive $v_{x}$,
the first (second) of Eqs. (\ref{EqC2}) is stable if we choose
the boundary condition at $-\infty $ ($ +\infty $). Using the
appropriate asymptotic condition one can find $a$ ($b$) at all
other points on the quasiclassical trajectory along the vector
${\bf v}_{F},$ integrating along the trajectory. Self-consistency
is achieved iteratively. The step-like approximation (\ref
{Eq2.3}) is used to find $a,b$ in the first approximation, which
are in turn substituted in Eq.\ (\ref{GapEq}) to find the next
iteration for $\Delta $. These steps are repeated until $\Delta$
does not change within numerical accuracy.

It is convenient to take the order parameter constant between
discrete points on the trajectory (the angle $\theta $ gives the
direction of $ {\bf v}_{F}$ as usual), separated by a distance
$h$. Then we find
\begin{eqnarray}
a_{i+1} &=&a_{i}+\frac{\Delta _{i}-a_{i}^{2}\Delta _{i}^{\ast
}-2a_{i}\omega}{a_{i}\Delta _{i}^{\ast }+\omega +\Omega _{i}\coth
(\Omega _{i}h/\cos
\theta )},  \nonumber \\
b_{i-1} &=&b_{i}+\frac{\Delta _{i}^{\ast }-b_{i}^{2}\Delta
_{i}-2b_{i}\omega }{b_{i}\Delta _{i}+\omega +\Omega _{i}\coth
(\Omega _{i}h/\cos \theta )} \label{EqC5}.
\end{eqnarray}
(We have explicitly taken into account that this procedure is
stable in opposite directions for $a,b$). Having obtained $a$ and
$b$, we can find $f$ and $g$ and therefore the order parameter
$\Delta$ and current density ${\bf j}$ using (\ref{GapEq}) and
(\ref{EqA6}).

\vspace{-2mm}
\subsection{Effect of surface reflectivity}

Our numerical method can also be applied when the transparency of
the junction is arbitrary, $0\leq D\leq 1$. Since part of the
quasiparticles get reflected from the boundary, the quasiparticle
trajectories of the reflected quasiparticles and the transmitted
ones from the other side will mix. Then even in the
non-self-consistent approximation for the order parameter
(\ref{Eq2.3}) we cannot simply impose continuity of the Green's
functions along a trajectory as we did before for the ideal
transparency case. Instead, one should use Zaitsev's boundary
conditions \cite{Zaitsev} which for $v_x>0$ is given by
\cite{BGZ95}
\begin{equation}
\widehat{d}^{~l}=\widehat{d}^{~r}\equiv \widehat{d}
\end{equation}
\begin{equation}
{\frac{D}{2-D}}\left[\left( 1+{\frac{\widehat{d}}{2}}\right)
\widehat{s}^{~r} ,\ \widehat{s}^{~l}\right] =\widehat{d}\
\widehat{s}^{~l2} \label{z}
\end{equation}
where
\begin{eqnarray}
&&\widehat{s}^{~r}=\widehat{G}_{\omega }^{r}({\bf
v}_{F},x{=}0)+\widehat{G}
_{\omega }^{r}({\bf v}_{F}^{\prime },x{=}0)  \nonumber \\
&&\widehat{d}^{~r}=\widehat{G}_{\omega }^{r}({\bf
v}_{F},x{=}0)-\widehat{G} _{\omega }^{r}({\bf v}_{F}^{\prime
},x{=}0)  \label{Eq2.16}
\end{eqnarray}
with ${\bf v}_{F}^{\prime }$ being the reflection of ${\bf
v}_{F}$ with respect to the boundary. Similar relations also hold
for $\widehat{s}^{~l}$ and $\widehat{d}^{~l}$. For $v_x<0$ one
has to replace $r \leftrightarrow l$ in (\ref{z}). In general, $D$
can be momentum dependent. In our calculations we consider
\cite{Yip94}
\begin{equation}
D(\theta )={\frac{D_{0}\cos ^{2}\theta }{1-D_{0}\sin ^{2}\theta
}}.
\end{equation}

Let us introduce four Green's functions
\begin{eqnarray}
&&\widehat{G}_{1}=\widehat{G}_{\omega }(0_-,\theta ),\quad
\widehat{G}_{3}=
\widehat{G}_{\omega }(0_-,\pi -\theta ) \\
&&\widehat{G}_{2}=\widehat{G}_{\omega }(0_+,\theta ),\quad
\widehat{G}_{4}= \widehat{G}_{\omega }(0_+,\pi -\theta ) \nonumber
\end{eqnarray}
We can now write
\begin{eqnarray}
&&\widehat{s}^{~l}=\widehat{G}_{1}+\widehat{G}_{3},\quad
\widehat{d}^{~l}=
\widehat{G}_{1}-\widehat{G}_{3}  \nonumber \\
&&\widehat{s}^{~r}=\widehat{G}_{2}+\widehat{G}_{4},\quad
\widehat{d}^{~r}= \widehat{G}_{2}-\widehat{G}_{4}
\end{eqnarray}
Rewriting these Green's functions in terms of $a_{\nu },b_{\nu }
\ (\nu =1,...,4)$, we can calculate $a_{1},b_{2},b_{3},$ and
$a_{4}$ by integrating from the corresponding infinity in the
direction of convergence [e.g.\ using Eq.\ (\ref{EqC5})]. At the
boundary, the remaining functions $b_{1},a_{2},a_{3},$ and
$b_{4}$ are obtained from Zaitsev's boundary conditions; the
remarkable results obtained by Eschrig \cite{Eschrig} are
\begin{eqnarray*}
b_{1} &=&\frac{Db_{2}+Rb_{3}+a_{4}b_{2}b_{3}}{1+a_{4}
\left(Db_{3}+Rb_{2}\right) }, \ a_{2}=\frac{Da_{1} + Ra_{4}
+ a_{1}a_{4}b_{3}}{1+b_{3}\left( Da_{4}+Ra_{1}\right) }, \\
a_{3} &=& \frac{Da_{4}+Ra_{1}+a_{1}a_{4}b_{2}}{1+b_{2}\left(
Da_{1}+Ra_{4}\right) }, \
b_{4}=\frac{Db_{3}+Rb_{2}+a_{1}b_{2}b_{3}}{ 1+a_{1}\left(
Db_{2}+Rb_{3}\right) }.
\end{eqnarray*}
Notice that $D=1$ gives: $a_1=a_2$, $a_3=a_4$, and $D=0$ gives:
$a_1=a_3$, $a_2=a_4$, which correspond to the completely
transparent and completely reflective cases respectively (the
same argument also holds for the $b$'s). The values of the $a$'s
and $b$'s at other points are calculated using Eq.\ (\ref{EqC5}).

\vspace{-2mm}
\subsection{Effect of surface roughness}

The effects of surface roughness are accounted for by introducing
a thin layer of impurities (elastic scatterers) of width $d$
\cite{rainer,OOM,GY}. In the Born approximation, the Eilenberger
equations in the layer are written as
\begin{eqnarray}
{\bf v}_{F}\cdot \frac{\partial }{\partial {\bf r}}
\hat{G}_{\omega }({\bf v}_{F},{\bf r}) &+& \left[
\omega_{_R}\hat{\tau}_{3}+\widehat{\Delta}_{_R}({\bf v}_{F},{\bf
r),}\hat{G}_{\omega }({\bf v}_{F},
{\bf r)}\right] =0 ; \nonumber \\
\widehat{\Delta}_{_R} &=& \left(
\begin{tabular}{ll}
0 & $\Delta_{_R}$ \\
$\Delta_{_R}^{\ast }$ & 0
\end{tabular}
\right) ,
\end{eqnarray}
where $$\omega_{_R}(x)=\omega +\frac{\left\langle g_{\omega }({\bf
v} _{F},x{\bf )}\right\rangle _{{\bf v}_{F}}}{2\tau },\
\Delta_{_R} (x)=\Delta +\frac{\left\langle f_{\omega }({\bf
v}_{F},x{\bf )} \right\rangle _{{\bf v}_{F}}}{2\tau },$$ $\tau
=v_{F}l$, and $l$ is the mean free path inside the layer. The
degree of roughness is given by the ratio $\rho =d/l$ in the
limit when $d\rightarrow 0,l\rightarrow 0$ simultaneously. For
strong scattering, the $x$-independent terms in the above
expressions can be dropped to obtain the Schopohl-Maki
transformed equations in the form
\begin{eqnarray}
-\cos \theta \frac{\partial a}{\partial \tilde{x} }
&=&2\tilde{\omega}a-\tilde{\Delta}+\tilde{\Delta}^{\ast }a^{2}; \\
\cos \theta \frac{\partial b}{\partial \tilde{x} }
&=&2\tilde{\omega}b-\tilde{\Delta}^{\ast }+\tilde{\Delta}b^{2},
\nonumber
\end{eqnarray}
where $\tilde{x} =x/d$ and
\begin{eqnarray*} \tilde{\omega}(\tilde{x})=\frac{1}{2}\rho \left\langle
g_{\omega }({\bf v}_{F},\tilde{x}d)\right\rangle _{{\bf v}_{F}},\
\ \tilde{\Delta}(\tilde{x})= \frac{1}{2}\rho \left\langle
f_{\omega }({\bf v}_{F},\tilde{x}d) \right\rangle _{{\bf v}_{F}}.
\end{eqnarray*}
In the above mentioned limit, integrating the equations over
$\tilde{x}$ from 0 to 1, and assuming that $f_{\omega },g_{\omega
}$ are slow functions of $x$, we can use Eq.\ (\ref{EqC5}) with
$h=1$ to calculate the jump of $a_{\nu },b_{\nu }$ across the
boundary. This approach also works for a rough surface (we put
formally $D=0$).

\vspace{-2mm}
\section{Order parameters in a $d$-wave superconductor}
\label{OP}

As mentioned in appendix \ref{QCEEq.}, the order parameter in a
superconductor is related to the anomalous Green's function via
the self-consistency equation (\ref{GapEq}). In general the
interaction potential $V({\bf v}_{F},{\bf v}'_{F})$ in
(\ref{GapEq}) can have components in different channels. Keeping
only the important terms, we can write
\begin{equation}
V({\bf v}_{F},{\bf v}_{F}^{\prime })=V_{d}\cos 2\theta \cos
2\theta' + V_{d'} \sin 2\theta \sin 2\theta' + V_s
\end{equation}
where $V_{d}$, $V_{d'}$, and $V_s$ are the components of the
interaction potential in the $d_{x^2-y^2}$, $d_{xy}$, and $s$
channels respectively. (In these expressions we assume
$x\parallel {\bf a}$ and $y\parallel {\bf b}$, since the
orientation of the order parameter is linked to the local crystal
axes of the system ${\bf a,b}$.) The order parameter in this case
can also have mixed symmetry
\begin{equation}
\Delta({\bf v}_F,{\bf r}) = \Delta_{d}({\bf r}) \cos 2\theta +
\Delta_{d'}({\bf r}) \sin 2\theta + \Delta_s ({\bf r})
\end{equation}
Substituting into (\ref{GapEq}) we find
\begin{eqnarray}
\Delta_{d}({\bf r}) &=& 4\pi \lambda_{d} T\sum\limits_{\omega
>0} \int
{d\theta \over 2\pi} f_\omega (\theta,{\bf r}) \cos 2\theta \\
\Delta_{d'}({\bf r}) &=& 4\pi \lambda_{d'} T\sum\limits_{\omega
>0} \int
{d\theta \over 2\pi} f_\omega (\theta,{\bf r}) \sin 2\theta \\
\Delta_s({\bf r}) &=& 2\pi \lambda_s T\sum\limits_{\omega >0} \int
{d\theta \over 2\pi} f_\omega (\theta,{\bf r})
\end{eqnarray}
where $\lambda_{d,d'}=N(0) V_{d,d'}/2$ and $\lambda_s=N(0) V_s$
are dimensionless interaction constants. These equations are used
in the self-consistent calculation of the components of the order
parameter.

In a homogeneous superconductor, the anomalous Green's function
$f_\omega$ takes the simple form
\begin{equation}
f_\omega (\theta) = \frac{\Delta (\theta)}{\sqrt{\omega
^{2}+\left| \Delta (\theta) \right| ^{2}}}
\end{equation}
Keeping only the $d_{x^2-y^2}$ component of the order parameter we
find

\begin{equation}
\Delta _d(T)=\lambda _{d} 4\pi T\sum\limits_{\omega >0}^{\omega
_{c}}\int\limits_{0}^{2\pi }\frac{d\theta }{2\pi }\frac{\Delta
_d(T)\cos ^{2}2\theta }{\sqrt{\omega ^{2}+\Delta _d(T)^{2}\cos
^{2}2\theta }} \label{gapeq2}
\end{equation}
where $\omega _{c}$ is the cutoff frequency. In our numerical
calculations we take $\omega_c=10 \pi$. At $T=0$, the right hand
side of Eq.\ (\ref{gapeq2}) diverges as $\Delta_d \rightarrow 0$.
Thus, no matter how small $\lambda_d$ is, there exists a finite
value for $\Delta_d$ that satisfies Eq.\ (\ref{gapeq2}). There
also exists a finite temperature $T_c$, below which $\Delta_d$ is
nonzero. This is not true for the subdominant order parameters,
because the presence of the dominant order parameter, at $T<T_c$,
introduces a cutoff which removes the divergence. As a result,
even below $T_{c2}$, the bulk of the superconductor contains only
one symmetry of the order parameter (the dominant one) even if
the interaction potential supports other symmetries as
well\cite{LLIX}. Near a (110)-surface of a $d$-wave superconductor
on the other hand, the dominant order parameter is suppressed and
subdominant gaps may appear. The resulting mixed order parameter
(e.g.\ $d+id'$ or $d+is$) breaks time reversal symmetry and
results in spontaneous currents near the surface. This behavior
is clearly demonstrated in Fig.\ \ref{fig14}a.

One can show that $\lambda_d$ is related to $T_c$ by
\begin{equation}
\lambda_d^{-1}= \ln {T \over T_c} + 2 \pi T \sum_{\omega
>0}^{\omega_c} {1 \over \omega} \label{lambda}
\end{equation}
The same equation holds for other symmetries of the order
parameter. In our numerical calculations we use (\ref{lambda}) to
find $\lambda_{d'}$ and $\lambda_s$ from $T_{c2}$ and $T_{cs}$
respectively.

At $T=0$, and in the weak coupling limit $\lambda _{d}\ll 1,$ it
follows from (\ref{gapeq2}) that

\begin{equation}
\Delta _d(0)=2\omega _{c}\beta e^{-1/\lambda _{d}},\quad \ln
\beta =\ln 2-1/2\approx 1.21.
\end{equation}
The critical temperature $T_{c}$ on the other hand is

\begin{figure}[t]
\epsfysize 4cm \epsfbox[10 150 400 410]{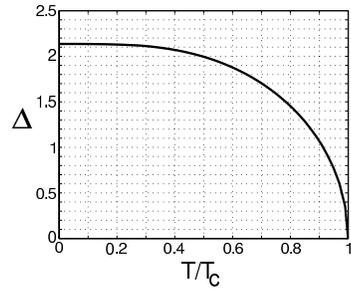}
\caption{Superconducting gap as a function of temperature.}
\label{fig17}
\end{figure}

\begin{equation}
T_{c}=\frac{2}{\pi }\omega _{c}\gamma e^{-1/\lambda _{d}},\quad
\ln \gamma=0.577,\ \gamma \approx 1.78.
\end{equation}
Thus, $\Delta _d(0)/T_{c}= \pi \beta /\gamma \approx 2.14$. In
terms of $T_{c}$, Eq.\ (\ref{gapeq2}) can be presented in the form
:

\begin{eqnarray}
\ln \frac{T}{T_{c}}=2\pi T\sum\limits_{\omega >0}^{\infty }\left(
2\int\limits_{0}^{2\pi }\frac{d\theta }{2\pi }\frac{\cos
^{2}2\theta }{\sqrt{\omega ^{2}+\Delta _d(T)^{2}\cos ^{2}2\theta
}}-\frac{1}{\omega }\right) \label{lnT} \nonumber
\end{eqnarray}
In the limiting cases, the solution of this equation reads

\begin{eqnarray}
\Delta _d(T)= \left\{
\begin{array}{ll}
\Delta _d(0) \left[ 1-3\zeta (3)\right] \left( \frac{T}{\Delta
_d(0)}\right) ^{3}, & T \ll T_{c} \\
\left( \frac{32\pi ^{2}}{21\zeta (3)}\right)
^{1/2}T_{c}(1-\frac{T}{T_{c}})^{1/2}, & T \lesssim T_{c}
\end{array} \right. \nonumber
\end{eqnarray}
For arbitrary temperatures $0\leq T\leq T_{c}$ the numerical
solution of Eq.\ (\ref{gapeq2}) is shown in Fig.\ \ref{fig17}.

\end{document}